\documentclass[twocolumn,amsmath]{aastex702}
\newcommand{\me}{\mathrm{e}}   
\newcommand{\mi}{\mathrm{i}}   
\newcommand{\sinc}{{\rm sinc}} 
\newcommand{\cs}{\texttt{CoherentSearch.jl}}
\newcommand{\riptide}{\texttt{riptide}}
\newcommand{\rseek}{\texttt{rseek}}
\newcommand{\accelsearch}{\texttt{accelsearch}}
\newcommand{\prepfold}{\texttt{prepfold}}
\newcommand{\rednoise}{\texttt{rednoise}}

\usepackage{algpseudocode}
\algrenewcommand\algorithmicindent{0.9em}
\algtext*{EndFor}                 
\algtext*{EndIf}                  

\shorttitle{A Coherent Harmonic Summing Pulsar Search}
\shortauthors{Ransom}
\graphicspath{{./}{figures/}}

\begin{document}

\title{A Coherent Harmonic Summing Pulsar Search Code}

\author[orcid=0000-0001-5799-9714,sname=Ransom,gname=Scott]{Scott M.~Ransom}
\affiliation{National Radio Astronomy Observatory, 520 Edgemont Road, Charlottesville, VA 22903, USA}
\email[show]{sransom@nrao.edu}

\begin{abstract}

Pulsars have narrow pulses, so their signals spread over many harmonics in a Fourier transform. Standard pulsar searches add the powers of those harmonics and discard their phases, which throws away the pulse shape and some of the signal. Folding the time series at each trial period keeps both, but is expensive even with the Fast Folding Algorithm (FFA). We describe \cs, an open-source Julia code that reconstructs the pulse profile at every trial spin frequency directly from the Fourier transform. It interpolates the complex amplitudes of typically 60 harmonics, inverse transforms them into a profile, and tests that profile against boxcar templates of many widths. Because the input spectrum is normalized, the noise in every profile is known in closed form, so one threshold gives one false-alarm rate across the whole search. In an injection study of over 860,000 simulated pulsars at a matched false-alarm rate, the new search recovers 76\% of the pulsars in white noise, against 64\% for \riptide's FFA in its recommended configuration and 42\% for PRESTO's \accelsearch, and its threshold does not move under red noise. On one CPU core it is 1.4--2.2 times faster than \riptide\ at matched frequency coverage, it scales to 27 times faster on 48 cores, and modern GPUs run it up to nine times faster than a 20-core workstation.

\end{abstract}



\section{Introduction} \label{sec:intro}

Most pulsars emit narrow pulses. The fraction of a rotation during which the pulse is ``on'', called the duty cycle, is typically a few percent, and for some pulsars well under one percent \citep{lk12,pkj+21}. A narrow periodic pulse train in a long time series has a Fourier transform with many significant components we call harmonics. In that long Fourier transform, the modulation theorem places the amplitudes of the Fourier transform of an individual pulse at integer harmonic multiples $hf$ of the fundamental spin frequency $f$. For narrow pulses, the harmonics above the fundamental carry most of the pulsed power.

Because of this, pulsar searches, which usually use the fast Fourier transform (FFT) for efficiency, add together the powers of many harmonics of each trial spin frequency to improve their sensitivity \citep{bc69,rem02,lk12}. This is called incoherent harmonic summing. ``Incoherent'' means that the phases of the complex Fourier amplitudes are discarded before summing, so the pulse shape is lost. It has been a standard of pulsar searching since the earliest digital searches \citep{bc69} since it improves search sensitivity and is relatively computationally efficient. Fourier phases \emph{have} been used to build rough pulse profiles of candidates by inverting the complex Fourier coefficients at a candidate frequency, but only for candidates already selected by an incoherent sum \citep{nft95}; PRESTO's \texttt{fourier\_fold.py} does the same today \citep{presto_ascl}.

We can do better if we use the phase of each harmonic in the search itself. Folding the time series at the spin frequency into $n$ phase bins, with $n$ large enough to resolve the pulse, is the familiar example: the folded profile keeps every harmonic, with its phase. But folding at each of $\sim N$ independent trial frequencies costs $O(N^2)$ operations for a time series of $N$ samples, since each sample has to be placed in a profile bin for every trial, and that has kept folding searches out of reach for blind surveys.

The Fast Folding Algorithm \citep[FFA;][]{staelin69} reuses partial sums recursively, like an FFT does, and brings the cost of a folding search to $O(N \log N)$. Until recently it was rarely used in real searches since implementations are intricate, and the constant in front of $N \log N$ was too large compared with FFT searches to make it worthwhile \citep{cbc+17,pkr+18}. The \riptide\ package \citep{mbs+20,riptide_ascl}\footnote{\url{https://github.com/v-morello/riptide}} has changed that. It is a fast and easy to use FFA implementation, and \citet{mbs+20} showed analytically that a folding search closely approaches the optimal sensitivity to any periodic signal and is more sensitive than the incoherent harmonic sum at every period and duty cycle. It is still much slower than FFT-based searches, but it has enabled more sensitive searches for realistic pulse profiles \citep[e.g.~][]{srp+22,wcb+24,gbm24}.

\citet{mbs+20} also noted that a search in the Fourier domain can reach the same sensitivity as the FFA if it sums harmonics \emph{coherently}, i.e. keeps their complex amplitudes and phases, and they suggested that such a search could be added to an existing FFT pipeline. This paper describes such a code. \cs\ \citep{coherentsearch_zenodo}\footnote{\url{https://github.com/scottransom/CoherentSearch.jl}} is a pulsar search written in Julia \citep{julia17} that takes the long FFT of a time series as its input, uses Fourier interpolation to obtain the complex amplitude of each harmonic at each trial frequency, and inverse transforms them into a pulse profile that it then tests for pulsar-like pulsations. The harmonics all come from one $O(N \log N)$ transform. Other efficiencies that are natural in the Fourier domain but awkward in the FFA's recursion give it a smaller constant in front of that cost, improved sensitivity, a calculable false-alarm rate, resilience to red noise, and a straightforward path to many-core CPUs and to GPUs. We describe the algorithm in \S\ref{sec:algorithm}, compare the code against \riptide\ and PRESTO's \accelsearch\ in a large injection study in \S\ref{sec:comparisons}, and conclude in \S\ref{sec:conclusions}.

\section{The Coherent Search Algorithm} \label{sec:algorithm}

The goal of the algorithm is to produce a pulse profile as a function of trial spin frequency $f$, and then to ask two things of each profile: is there a significant pulsation, and is it pulsar-like? An incoherent harmonic sum can only answer the first. It adds power from whatever is at the harmonic frequencies, whether that is a narrow pulse, a sinusoid, or a chance alignment of noise and interference, and it cannot tell them apart.

We start with a normalized FFT of a long, evenly sampled time series (\S\ref{sec:input}) which we step finely through at a grid of Fourier frequencies $r = fT$, where $T$ is the length of the observation, covering the range of spin frequencies we want to search. At each $r$ we determine the complex Fourier amplitudes of each of $H$ harmonics $hr$ (\S\ref{sec:fourierinterp}), and inverse FFT them into a real-valued pulse profile. For each profile we compute a detection statistic for both significance and whether the pulse is pulsar-like (\S\ref{sec:metric}). Finally, we reuse subsets of the same $H$ amplitudes to build lower-resolution profiles at integer multiples of $r$, and test those with the same statistic (\S\ref{sec:decimation}).

\subsection{The Algorithm in Pseudocode} \label{sec:pseudocode}

Figure~\ref{fig:pseudocode} shows the whole search. The input, $F$, is an FFT of a long, evenly sampled time series whose Fourier amplitudes have been scaled to have unit mean. The search walks a grid of trial fundamental Fourier frequencies $r$ (line 1). At each trial it interpolates the complex Fourier amplitude of each of the $H$ harmonics of that fundamental (lines 2--3), stacks those amplitudes into a half-complex spectrum with the DC (i.e.~zero-frequency) term held at zero, and inverse transforms the stack into a real pulse profile of $n = 2H$ phase bins (lines 6--7). Because the DC term is zero, the mean of every profile is exactly zero. The profile is then correlated against a bank of boxcars of different widths and at every phase, and the largest of those correlations is the trial's signal-to-noise ratio (lines 9--14). Trials scoring above the threshold become candidates.

Lines 4--7 compute the higher spin frequencies almost for free. A signal at $k$ times the trial fundamental has its harmonics at Fourier frequencies $kr, 2kr, \ldots$, which are amplitudes the search has already interpolated as harmonics $k, 2k, \ldots$ of $r$. Taking every $k$th amplitude therefore gives the harmonic stack of the multiple at the cost of one short inverse FFT and no interpolation at all. We describe this harmonic decimation in \S\ref{sec:decimation}.

We distribute a literal implementation of Figure~\ref{fig:pseudocode} alongside the production code, as \texttt{bin/toy\_coherent\_search.jl}. Its functions carry the line numbers of the figure, it uses brute-force interpolation, one inverse FFT per fold, and the boxcar filter evaluated straight from its definition. It is checked against the production search via the test suite. It is a complete and working search, but it is 150--250 times slower than the production code described below, so it is meant to be read rather than run.

\begin{figure}[t]
\rule{\linewidth}{0.8pt}\vspace{2pt}
{\footnotesize
\begin{algorithmic}[1]
\For{$r = r_{\rm lo},\ r_{\rm lo}+\delta r,\ \ldots,\ r_{\rm hi}$}\Comment{Fourier frequency trials}
  \For{$h = 1, \ldots, H$}\Comment{Determine harmonics}
    \State $A_h \gets \mathrm{interpolate}(F,\, h\,r,\, m)$\Comment{Fourier interpolation}
  \EndFor
  \For{$k = 1, \ldots, K$}\Comment{Decimation factors}
    \State $H_k \gets \lfloor H/k \rfloor$, \ \ $n \gets 2 H_k$\Comment{\# harmonics, profile bins}
    \State $S \gets (0,\, A_{k},\, A_{2k},\, \ldots,\, A_{H_k k})$\Comment{Stack harmonics}
    \State $P \gets \mathrm{inverseRealFFT}(S)$\Comment{Reconstruct profile}
    \State $\sigma \gets \mathrm{noiseScale}(P)$, \ \ $S_{\rm tot} \gets \sum_j P_j$\Comment{Profile stats}
    \State $b \gets -\infty$\Comment{Initialize best S/N}
    \For{each boxcar width $w$, \ $\delta \gets w/n$}\Comment{Boxcar filters}
      \For{$p = 0, \ldots, n-1$}\Comment{Boxcar phase trials}
        \State $S_w \gets \sum_{i=0}^{w-1} P_{(p+i) \bmod n}$\Comment{Boxcar signal sum}
        \State $s \gets (S_w - \delta\,S_{\rm tot}) \big/ \big(\sigma\sqrt{w\,(1-\delta)}\,\big)$\Comment{S/N statistic}
        \State $b \gets \max(b,\, s)$\Comment{Update best S/N}
      \EndFor
    \EndFor
    \If{$b >$ threshold}
      \State record a candidate at $k\,r/T$ with $H_k$ harmonics
    \EndIf
  \EndFor
\EndFor
\State collapse near-duplicate, then harmonically related, candidates
\State report the strongest candidates
\end{algorithmic}
}
\vspace{2pt}\rule{\linewidth}{0.8pt}
\caption{The coherent harmonic-summing search. $F$ is the normalized Fourier amplitude array (i.e.~a complex FFT) of an observation of duration $T$; $H$ is the number of harmonics, $m$ the number of neighboring FFT bins used for the interpolation, $K$ the largest harmonic decimation factor, and $r_{\rm lo} = f_{\rm lo}T$ and $r_{\rm hi} = f_{\rm hi}T$ bound the fundamental frequencies searched, in steps of $\delta r = 1/(2H)$. Lines 1--16 produce one pulse profile and one signal-to-noise ratio for every trial fundamental and every decimation factor; lines 17--18 are candidate bookkeeping. Ends of blocks are marked by indentation alone, so that these line numbers are also the line numbers of the reference implementation distributed with the code, \texttt{bin/toy\_coherent\_search.jl}. The production code computes the same quantities but reorganized and heavily optimized for speed.}
\label{fig:pseudocode}
\end{figure}

In the subsections below we describe the details of the Julia implementation that make the production code fast, sensitive, and parallel.

\subsection{Code basics} \label{sec:code}

The code is written in Julia \citep{julia17}. Julia compiles to native machine code ahead of use, so numerical loops run at the speed of C or Fortran while the source stays readable. It has built-in support for the three kinds of parallelism this search uses: SIMD instructions (i.e. one instruction operating on several numbers at once) within a core, threads across the cores of a machine, and GPUs through the \texttt{CUDA.jl} package \citep{cudajl18}. Installing the code and its dependencies is one command, and it needs no compiler or system libraries.

We worked out the algorithm and tested various aspects of it with a reference implementation, \texttt{coherent\_search}\footnote{\url{https://github.com/scottransom/coherent_search}}, which we wrote first, using Python and NumPy. The Julia code is validated against it, such that interpolated amplitudes, reconstructed profiles, and the detection statistic all agree with the Python code to $\sim 10^{-16}$ relative precision. Inside the Julia package, a deliberately unoptimized reference path is kept alongside the optimized search, and the test suite verifies that the two agree to $\sim 10^{-15}$ on real data. Every optimization has to keep both comparisons intact. The test suite also checks the analytic noise scale of \S\ref{sec:input} against simulated noise, the decimated folds of \S\ref{sec:decimation} against direct folds at the multiplied frequency, and the GPU search against the CPU search. The code has been run on simulated pulsars and on real Parkes and Green Bank Telescope search data. All of the tests, benchmarks, and simulations in this paper used version 1.0.0 of the code or earlier commits with identical search code.

\subsection{How many profile bins do we need?} \label{sec:nbins}

The number of harmonics $H$ sets the number of profile bins $n = 2H$, and with it the narrowest pulse the search can resolve. Canonical (i.e.~non-recycled) pulsars have narrow duty cycles, and the duty cycle correlates with period. In the MeerKAT Thousand-Pulsar-Array sample of \citet{pkj+21}, the median full-width-at-half-maximum duty cycle is 1.9\% at a period of one second, falling as $P^{-0.35}$, and 12.6\% of the slow pulsars have duty cycles below 1\%, with 2.1\% below 0.5\% \citep[see also][]{lk12}. Millisecond pulsars have broader pulses, typically 5--30\% of a rotation.

We set $H = 60$ by default, i.e.~120 profile bins, which resolves a duty cycle of just under 1\% with one bin across the pulse. That is enough for most of the pulsar population. Pulses narrower than that lose sensitivity to harmonic truncation, and we return to them in \S\ref{sec:sensitivity}. Wider pulses need far fewer harmonics, and searching them with fewer is a key efficiency that we describe in \S\ref{sec:decimation}. We choose $H = 60$ because it is divisible by 2, 3, 4, 5, and 6, which is what the decimation needs.

\subsection{Preparing the input FFT} \label{sec:input}

The search reads a PRESTO-format FFT of a dedispersed time series, produced by \texttt{realfft} and then normalized by PRESTO's \rednoise\ \citep{presto_ascl}. \rednoise\ divides each complex amplitude by the square root of the median power in a block of neighboring frequency bins, using blocks that grow from a few bins at the lowest frequencies to $\sim 100$ bins above 6\,Hz. The result is a spectrum in which the mean noise power is one at every frequency. It is both whitened, such that the slow ``red'' noise from real-world effects such as gain variations and atmospheric fluctuations is flattened, and normalized.

This matters for the search in two ways. First, we can analytically compute the noise statistics of everything that happens later in the algorithm. A mean Fourier power of one means that the real and imaginary parts of each amplitude have variance of one half. The unnormalized inverse FFT of a stack of $H$ harmonics with the DC term set to zero then has, in every profile bin, a noise standard deviation of
\begin{equation} \label{eq:sigma}
     \sigma = \sqrt{2\,n_{\rm low} + \tfrac{1}{2}\,n_{\rm nyq}},
\end{equation}
where $n_{\rm low}$ counts the harmonics below the profile's own Nyquist bin and $n_{\rm nyq}$ is one if that bin carries data. This is $\sqrt{n}$ times a correction $\sqrt{1 - 3/(4H)}$ that is 0.6\% at $H = 60$ but 3.8\% at $H = 10$. Without it, the shallow (i.e.~fewer harmonics) folds of \S\ref{sec:decimation} would be statistically biased against the deep (i.e.~more harmonics) ones. We compute $\sigma$ rather than estimate it from each profile, which turns out to be both faster and better. An estimate from a profile of $n$ bins carries a sampling error of order $1/\sqrt{n}$, and the reported signal-to-noise ratio is exactly proportional to $1/\sigma$, so that error would apply to every candidate. Measured over many trials, the analytic $\sigma$ tracks the exact noise level to about 1\%, better than the robust estimator it replaced, and it removes a source of variance from the statistic that we return to in \S\ref{sec:fap}. The option \texttt{--sigma measured} estimates $\sigma$ from the data instead, which is the right choice when the noise level varies with frequency, for example because of residual red noise or strong frequency-related sets of interference signals. The search scores a few groups of trials both ways and warns if the two disagree by more than 10\%, because an unnormalized input inflates every signal-to-noise ratio silently.

Second, we set the DC (i.e. zero-frequency) amplitude to zero explicitly, so the mean of every reconstructed profile is identically zero. There is then no baseline to estimate in each reconstructed profile. That removes a term from the variance of the detection statistic and, as \S\ref{sec:metric} shows, simplifies its calculation.

\subsection{Fourier Interpolation} \label{sec:fourierinterp}

An FFT gives the Fourier amplitudes only at integer Fourier frequencies $r_k = k$, corresponding to a spacing of $1/T$ in Hz. A pulsar's harmonics almost never fall on those integers, and between two adjacent bins the phase of a harmonic rotates through a full turn. To fold coherently we need the amplitude and phase of every harmonic at the trial frequency itself, to a precision comparable to the duty cycle we are searching for. So a 1\% duty cycle needs harmonic phases accurate to a few degrees. Reading the nearest bin achieves nowhere near that phase accuracy. We also lose amplitude when a signal sits between bins, an effect known as scalloping that reaches 36\% in power halfway between bins \citep{rem02}, and that the FFA, which folds directly in the time domain, does not suffer \citep{mbs+20}.

Fourier interpolation solves both problems. As described in \citet{mdk93,rem02}, the complex Fourier amplitude spectrum produced by an FFT can be interpolated at an arbitrary Fourier frequency $r = fT$ to provide the complex amplitude $A_r$ via
\begin{equation} \label{eq:interp}
     A_r \simeq \sum_{k=[r]-m/2}^{[r]+m/2} A_k \ \me^{-\mi \pi(r-k)} \sinc\left[\pi(r-k)\right],
\end{equation}
where $A_k$ are the complex Fourier amplitudes at the nearby integer Fourier frequencies $k$, $\sinc(x) = \sin(x)/x$ is the normalized sinc function, $[r]$ is the nearest integer to $r$, and $m$ is the total number of nearby FFT bins used for the interpolation. The sum is exact as $m \to \infty$ and truncating it at $m$ terms leaves out the tails of the sinc response.

The truncation is well behaved. For a signal at fractional offset $\Delta r$ from a bin, the interpolated amplitude is the true one multiplied by a \emph{real} factor $S_m = \sum_k \sinc^2[\pi(\Delta r - k)]$ over the $m$ terms kept, so truncation introduces no phase error at all, only a loss of amplitude. The lost fraction of signal power is $1 - S_m \simeq 4\sin^2(\pi\Delta r)/(\pi^2 m)$, or $\sim$$0.2/m$ averaged over offsets and $0.4/m$ at worst. The noise is truncated in the same proportion, so the recovered signal-to-noise ratio scales as $\sqrt{S_m}$. Our default is $m = 16$, which loses 1.3\% of the signal power on average, against 0.6\% at $m = 32$ for twice the cost. The loss per harmonic does not depend on $H$, so a search with more harmonics does not need a larger $m$; if anything the sum over harmonics averages the offset-dependent part of the loss down further.

The other choice is how finely to step $r$. Harmonic $h$ moves by $h\,\delta r$ per step in the fundamental, so the highest harmonic $H$ sets the step. We step so that harmonic $H$ advances by half a Fourier bin, corresponding to $\delta r = 1/(2H) = 1/n$. A signal then sits at most a quarter of a bin from a trial at its top harmonic, where the sinc response loses 19\% of the power, and 6.5\% averaged over offsets, whereas lower harmonics sit proportionally closer to their trials and lose less. This is the same rule the FFA uses. \riptide\ folds a base period of $b$ samples into $b$ bins and produces trials spaced by $1/b$ Fourier bins, one phase bin of drift over the observation \citep{mbs+20}, and we have verified against \riptide\ itself that its trial spacing equals ours at every fold depth to $2\times 10^{-4}$.

Two properties of the problem make Equation~(\ref{eq:interp}) cheap to evaluate exactly at every trial. The coefficients factor: $\me^{-\mi\pi(r-k)}\sinc[\pi(r-k)] = A(\Delta r)/(\Delta r - j)$ with $j$ the integer offset of the bin, where $A(\Delta r)$ is one complex scalar per evaluation point and $1/(\Delta r - j)$ is real. One interpolated amplitude therefore costs $m$ real-times-complex multiply-adds and reads only $m$ consecutive bins of the FFT. And because the trial grid is rational, with $\delta r = 1/n$, only $n$ distinct fractional offsets $\Delta r$ occur for any harmonic in the whole search. The weights are tabulated once per harmonic and referenced by an integer index that advances by a fixed amount per trial, so the inner loop has no divisions and no transcendental functions. The FFT-correlation method that the Python reference implementation uses \citep[the ``fine grid'' of][]{rem02} is slower here by a factor of several because most of the fine-grid points it computes are never used, and it is less accurate ($\sim 10^{-2}$ in amplitude against $\sim 10^{-7}$) because it linearly interpolates between them.

Each trial $r$ is computationally independent of every other, so trials can be distributed across CPU cores and GPU threads freely. But consecutive trials read overlapping windows of the same $m$ bins per harmonic, which makes the interpolation fast in practice. A group of $V$ consecutive trials becomes a small matrix-vector product between a tabulated $V \times (m + \Delta)$ weight block and one contiguous run of FFT bins, where $\Delta$ is the drift of the window across the group. Computationally, each lane of a SIMD register then computes a different trial, with no gathers and no cross-lane reductions. On the CPU this is done with $V = 32$ trials at a time. On the GPU each thread takes one trial and a block of threads reads the shared bin window once. Trials are also processed in groups of a few thousand, so that each harmonic's window of FFT bins is loaded into cache once and reused by every trial in the group (\S\ref{sec:details}).

\subsection{Harmonic Decimation} \label{sec:decimation}

One of the most expensive parts of the search is the interpolation of $H$ harmonics at each trial. Once we have them, we have at least some of the harmonics of higher spin frequencies as well. A signal at $k$ times the trial fundamental $r$ has its own harmonics at Fourier frequencies $kr, 2kr, 3kr, \ldots$, and those are exactly the amplitudes already computed for harmonics $k, 2k, 3k, \ldots$ of $r$. Selecting every $k$th amplitude gives the harmonic stack of a signal at $kr$ with $H_k = \lfloor H/k \rfloor$ harmonics, and one inverse FFT of length $H_k$ turns it into a profile with $n_k = 2H_k$ bins. We then apply the same boxcar statistic to it. We call this harmonic decimation.

\begin{figure*}[t]
\centering
\includegraphics[width=0.95\textwidth]{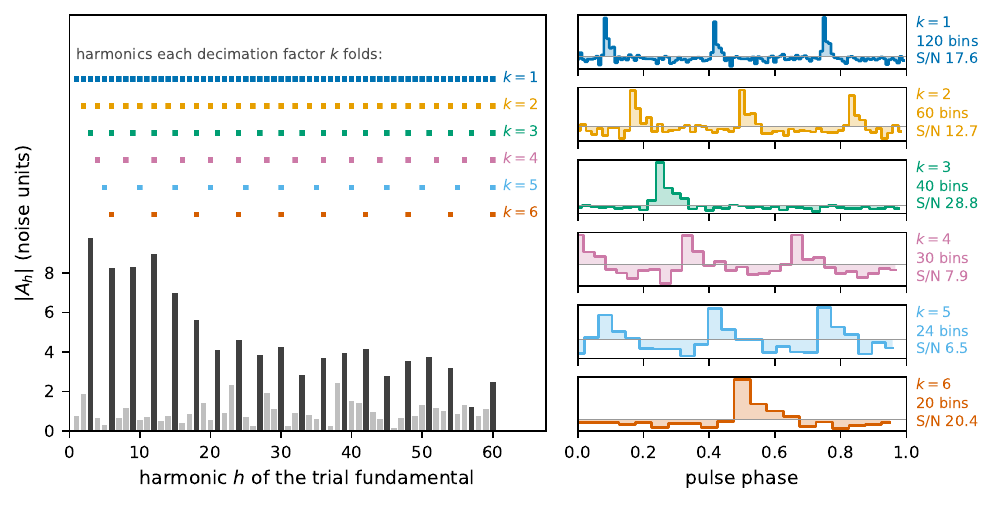}
\caption{Harmonic decimation, for a simulated pulsar with a narrow main peak and a weaker, wider shoulder, at three times the trial fundamental frequency. Left: the 60 interpolated harmonic amplitudes at the trial, in units of the noise, with the harmonics that carry signal in black; the rows of squares above mark the harmonics that each decimation factor $k$ folds. Right: the six profiles that those subsets produce, each in units of its own noise standard deviation, with the number of bins and the boxcar signal-to-noise ratio of each. The $k = 3$ fold recovers the pulse; the other folds place the same amplitudes at the wrong multiple of the period and look less like a pulsar. The signal was injected at a matched-filter signal-to-noise ratio of 32.}
\label{fig:decimation}
\end{figure*}

The cost is one short inverse FFT and one matched-filter pass per decimation factor, and no interpolation. With $H = 60$ and $k = 1, \ldots, 6$ the profiles have 120, 60, 40, 30, 24, and 20 bins, and each pass covers spin frequencies $k$ times those of the base grid. The decimated folds remain properly sampled since pass $k$ steps its fundamental by $k\,\delta r$, but its top harmonic is $H_k$, so that harmonic advances by $H_k k\,\delta r \le H \delta r$, never coarser than the base search. The read depth into the input FFT is also unchanged, since every decimated harmonic sits at a frequency the base fold already reads. So a default search of fundamentals from 0.1 to 125\,Hz reaches spin frequencies of 750\,Hz, past the fastest known pulsar, for the cost of interpolating to $60 \times 125$\,Hz.
Figure~\ref{fig:decimation} illustrates this with a simulated pulsar whose spin frequency is three times the trial fundamental. Only every third harmonic of the trial carries signal, so the profiles that fold every $k$th harmonic for $k \ne 3$ mix the pulsar's harmonics with noise and fold them at the wrong multiple of the period. The $k = 3$ fold is the pulsar, and it has the highest signal-to-noise ratio.

Decimation is more than a fast and efficient way to reach high frequencies. It is also the harmonic-summing ``ladder'' that FFT searches have always used, summing 1, 2, 4, 8, and 16 harmonics and keeping the best. A broad pulse has little power above its first few harmonics, and a fold that includes higher harmonics adds their noise and none of their signal. A fold of $H_k = 10$ harmonics into 20 bins is therefore the better detector of a 10\% duty cycle pulse, and it comes basically free from the searches of the lower frequencies, with more harmonics. The passes overlap deliberately such that every frequency below $f_{\rm hi}$ is folded at each of the six depths, and the best is kept. A model of the recovered signal-to-noise ratio for Gaussian pulses across the duty cycle distribution of \S\ref{sec:nbins} puts a single-depth search 8\% below the full ladder, which is our default.

Some of the resulting (fold depth, boxcar width) combinations are redundant. A boxcar of width $W$ bins on an $n$-bin profile weights harmonic $h$ by the Dirichlet kernel, which depends on $W$ and $n$ only through the duty cycle $\delta = W/n$. Since $n = 2H/k$, the pair $(k, 2W)$ and the pair $(2k, W)$ apply the same filter, differing only in where the harmonic sum truncates. A width of one bin on any but the deepest fold is the same filter as two bins one rung deeper at twice the resolution, and a wide boxcar on a shallow fold drags in harmonics that a broad pulse does not have. We therefore keep boxcar widths of 2--6 bins on every fold, plus one bin on the deepest fold and the full set of wide boxcars on the shallowest, corresponding to 25 of 38 combinations at the defaults. Modeled over Gaussian, scattered, two-component, and interpulse profiles at 60 duty cycles, the pruned set loses no sensitivity and saves 35\% of the matched-filter work.

\subsection{Calculation of the search metric} \label{sec:metric}

Ideally we would matched filter each profile with the true pulse shape, but in a search that shape is unknown. Following \citet{mbs+20} we use a bank of boxcars of increasing width instead. They are cheap to apply, their noise statistics are simple, and a boxcar of the right width recovers about 94\% of the matched-filter signal-to-noise ratio of a Gaussian pulse \citep{mbs+20}. The bank is geometrically spaced by a factor of 1.5, starting at one bin and stopping at 30\% of the profile, and is built from each profile's own $n$ so that every fold depth is filtered to the same range of duty cycles. \riptide\ instead builds one bank for each range of fold depths, from the smallest number of bins in that range. Where the range is wide, as in the 20 to 120 bins of configuration A (\S\ref{sec:comparisons}), its deepest folds reach only 5\% duty cycle. The narrow ranges of configurations B and C keep that limit between 20\% and 29\%.

For a profile $P_j$ with $n$ bins, a boxcar of width $w$ starting at phase bin $p$ has the sum $S_w(p) = \sum_{i=0}^{w-1} P_{(p+i) \bmod n}$. \riptide's statistic \citep[its \texttt{snr1}; the template is Eq.~A8 of][made zero-mean]{mbs+20} correlates the profile with the boxcar after making the template zero-mean and of unit norm, which gives
\begin{equation} \label{eq:snr1}
   {\rm S/N} = \max_{w,p}\ \frac{S_w(p) - \delta\, S_{\rm tot}}{\sigma\sqrt{w(1-\delta)}}, \qquad \delta = w/n,
\end{equation}
where $S_{\rm tot}$ is the sum of the whole profile and $\sigma$ the noise per bin. Every $(w, p)$ term is then a unit-variance Gaussian variable under white noise, so one threshold means one false-alarm probability per trial. We have verified our implementation of Equation~(\ref{eq:snr1}) against the \riptide\ binary on identical profiles to $10^{-7}$, its own single-precision rounding, and the two codes report the same quantity.

Our statistic differs from \riptide's in two ways. First, our profiles have zero mean by construction (\S\ref{sec:input}), so $S_{\rm tot} \equiv 0$ and the baseline term vanishes. \riptide\ removes the mean of the time series before folding, so the term is small there too, but in its case each profile's baseline still fluctuates. Second, and more important, the denominator of Equation~(\ref{eq:snr1}) assumes that the $n$ profile bins are independent. \riptide's are, since each bin of a time-domain fold sums a disjoint set of samples. Ours are not. A profile built from $H$ harmonics of a stack whose higher frequencies are zero is band-limited, and band-limited noise is correlated between neighboring bins. We therefore divide by the exact variance of the boxcar sum instead. Writing the boxcar's response at profile harmonic $j$ as the Dirichlet kernel $D_j = \sin(\pi j w/n)/\sin(\pi j/n)$, and with each filled harmonic contributing independent real and imaginary parts of variance one half,
\begin{equation} \label{eq:varsw}
   {\rm var}(S_w) = 2\sum_{j\ {\rm filled},\ j<n/2} |D_j|^2 \;+\; \tfrac{1}{2}\,|D_{n/2}|^2 ,
\end{equation}
with the last term present only when the profile's own Nyquist harmonic is filled. This was checked against $10^5$ simulated noise profiles at every combination of fold depth and fill count in the default search and agrees to the 0.2\% precision of the simulation. On a fully filled profile it differs from $\sigma\sqrt{w(1-\delta)}$ by a factor of about $1 + 3/(4n)$, 0.6\% at 120 bins but 4\% at 20, so without it the shallow folds of the ladder would win spuriously by that margin. Where harmonics are missing because they lie above the Nyquist frequency of the data (\S\ref{sec:details}) the difference is much larger.

The boxcar sums for all widths are computed from one prefix sum of the profile, $C_p = \sum_{i<p} P_i$ over two concatenated copies of the profile, as $S_w(p) = C_{p+w} - C_p$; this is the ``strided differences'' trick of \citet[][their \S5.4]{mbs+20}, and it makes each width cost one pass over the profile rather than $w$. The same filtering could be done in the Fourier domain by weighting the harmonic amplitudes before the inverse transform, and for a bank of Gaussian templates that would probably be the better route. For boxcars the prefix sum is hard to beat. In the code, the scan runs across 128 profiles at once so that its short inner loop, only 20--120 bins long, fills the SIMD lanes; a single-precision first pass discards the great majority of trials that cannot reach the threshold and a double-precision pass re-scores the rest.

None of the pieces of the statistic depend on the trial frequency or the fold depth. The input spectrum has a mean noise power of one at every frequency, Equation~(\ref{eq:sigma}) gives the noise in each profile bin from that normalization, and Equation~(\ref{eq:varsw}) gives the exact noise of every boxcar sum on every fold. Under white noise the statistic therefore has the same unit-variance distribution everywhere in the search, and one threshold gives the same false-alarm probability per trial at all frequencies and fold depths. Because $\sigma$ is computed rather than estimated, there is also no per-profile noise estimate to scatter the reported values. \S\ref{sec:fap} confirms both properties with the simulations, and shows that they are most of the reason the search can run at a lower threshold than the other codes at the same false-alarm rate.

\subsection{Other details} \label{sec:details}

\paragraph{Grouping the trials.} Trials are processed in groups of consecutive $r$ (2048 on the CPU, 65,536 on the GPU by default), and one group is handed to one thread or one GPU launch. Each harmonic's window of FFT bins, about 0.3\,kB for $h = 1$ up to 8\,kB for $h = 60$, is then read from main memory once per group and from cache by every trial in it. The group size changes nothing in the search itself, since the trial grid is always anchored to the global trial index. It substantially changes the speed, though, especially on GPUs. The best group size for a GPU depends on how its cache compares with its number of streaming multiprocessors, and on the eight cards we have measured it spans a factor of 128, from 8192 to 1,048,576 trials. Using the CPU's default on a GPU costs between 1.3 and 6.8 times the best speed for that card, with no change in what is searched. The code includes a script that determines the best group size on a new GPU card.

\paragraph{Harmonics above Nyquist.} For a trial at high frequency, harmonic $h$ may lie above the Nyquist frequency of the data. Those parts of the harmonic stack are left at zero and their number is recorded. The zeroed harmonics carry no signal and no noise, so both the noise scale of Equation~(\ref{eq:sigma}) and the boxcar variance of Equation~(\ref{eq:varsw}) count only the harmonics that are filled. Not accounting for this and using the nominal $H$ in Equation~(\ref{eq:sigma}) overestimates $\sigma$ and effectively suppresses fast candidates. Using $\sigma\sqrt{w(1-\delta)}$ in Equation~(\ref{eq:snr1}) underestimates the noise of a truncated fold by about $\sqrt{n/2H_{\rm filled}}$. We tested the correction on 500 simulated observations of normalized white noise, sampled at 60\,$\mu$s so that the 60th harmonic crosses the Nyquist frequency at a fundamental of 139\,Hz, and searched four 1\,Hz bands with 60 harmonics and no decimation. Without the correction, the highest statistic in each band averaged $5.31 \pm 0.23$ at 20\,Hz and $5.30 \pm 0.22$ at 100\,Hz, but $6.92 \pm 0.35$ at 271\,Hz and $8.90 \pm 0.49$ at 500\,Hz, and every one of the 500 observations had a noise peak above 6 in both of the high bands. With the correction, the four bands averaged $5.28 \pm 0.23$, $5.27 \pm 0.22$, $5.12 \pm 0.26$, and $4.91 \pm 0.26$, and a noise peak above 6 occurred in 0.2--1\% of the observations in every band. The slightly lower peaks at high frequency are likely because a fold with fewer harmonics has fewer independent trials.

\paragraph{Candidates.} A real signal will trigger a series of adjacent trials at every fold depth that covers it, and will also appear at $f/2$, $2f$, $3f/2$, and so on. Two passes through the candidate list remove these. Near-identical candidates within one Fourier bin are reduced to the strongest, and candidates whose frequencies form a ratio of small integers are reduced to the strongest member of the family. For each surviving candidate the code re-folds the profile using higher-precision ($m = 64$) Fourier interpolation at the full harmonic depth, reports its frequency, period, signal-to-noise ratio, harmonic count, and the duty cycle of the best-fitting boxcar, and can plot the profile if requested.

\section{Comparisons with existing Search Codes} \label{sec:comparisons}

In order to investigate how the new search performs against established pulsar search codes, used the way their authors intend, we performed comparisons against three of them. PRESTO's \accelsearch\ \citep{rem02,presto_ascl} with 16 harmonics and no acceleration search is a traditional incoherent harmonic summing search that would be used when a survey wants a fast search targeting canonical isolated pulsars. \riptide's \rseek\ \citep{mbs+20} is the state-of-the-art FFA. PRESTO's \prepfold, folding at the \emph{known} period of each injected pulsar, is more of a ``best case'' than a search competitor. Since \prepfold\ is given the known spin period, it pays no penalty for the number of trials a blind search must make, and therefore shows how much sensitivity a real search gives up.

\riptide\ needs several calling configurations, and the sections below specify which one each number refers to. \rseek\ covers a period range by downsampling the time series and folding into between $b_{\rm min}$ and $b_{\rm max}$ bins. Its documentation recommends a narrow range, with $b_{\rm max}$ about 10\% above $b_{\rm min}$, and its authors' pipeline searches several such ranges. Configuration C follows that recommendation, as suggested to us by V.~Morello. It folds into 20--120 bins only for periods shorter than 7.2\,ms (i.e.~120 samples), where no deeper fold fits, and into 120--140 bins from 7.2\,ms to 10\,s. Slow pulsars are therefore folded at about the depth of our own deepest folds, and with the same nine boxcar widths. Because it is \riptide\ in its recommended regime, configuration C is the one we compare against first. Configuration A uses $b_{\rm min} = 20$ and $b_{\rm max} = 120$ over the whole range of 1.33\,ms to 10\,s. That matches our default frequency coverage and fold depths exactly (60 harmonics and decimation by up to 6 give profiles of 120 down to 20 bins over the same range), and so allows us to compare the codes for an equivalent amount of work. Configuration B is the deeper search that \riptide's own pipeline uses and that its authors would choose for narrow pulses. It uses four period ranges, with the 20--120 profile bins of configuration A below 10\,ms, 160--174 bins from 10 to 50\,ms, 480--520 bins from 50 to 250\,ms, and 960--1040 bins above 250\,ms. It folds narrow slow pulsars six times more finely than we do. On one machine, configurations A, C, and B take roughly 1.7, 3, and 7 times our run time (\S\ref{sec:speed}). Configurations B and C are each run as separate \rseek\ calls over their period ranges, with the candidates combined under one threshold. The period boundaries of all three configurations are specific to our simulated data. They depend on the sampling time (60\,$\mu$s here) through the shortest period that fits a given number of bins, and the range of periods worth searching depends on the observation length (1006\,s here). A search of real data with a different sampling time or integration time needs its own ranges, which is part of what makes the FFA nontrivial to use well.

\subsection{Monte Carlo Simulations} \label{sec:mc}

We performed an extensive injection Monte Carlo experiment. Each simulated observation is $2^{24}$ samples at 60\,$\mu$s, i.e.~1006\,s of unit-variance Gaussian noise, into which we inject six fake pulsars. Every code then searches the identical data: \prepfold\ and \rseek\ read the time series, \accelsearch\ the FFT, and \cs\ the \rednoise-normalized FFT. We also ran \accelsearch\ on the normalized FFT, which changed its results by a few percent and is not shown. The study comprises 160,976 observations and 869,574 injected pulsars: 76,105 observations in white noise and 84,871 with red noise added, with 411,078 of the injections paired between the two. We later ran \rseek\ configuration C on 22,368 of the white-noise observations, and \rseek\ on whitened time series (\S\ref{sec:sensitivity}) on 33,010 of the red-noise observations, in each case together with every other code on the same observations. On those subsets the results for \cs\ agree with the full study to within 0.3\%, so the subsets are representative.

\paragraph{The injected population.} Two thirds of the injected pulsars are slow, with periods drawn log-uniformly from 20\,ms to 5\,s; one third are millisecond pulsars with periods drawn from a Gaussian of mean 3.5\,ms and width 1\,ms, truncated to 1.5--10\,ms. The slow pulsars take their duty cycles and profile shapes from the measured widths of the 1176 pulsars in the MeerKAT Thousand-Pulsar-Array sample \citep{pkj+21}, by bootstrapping that table's residuals about its duty-cycle versus period relation and its ratio of 10\% to 50\% widths. We kept scattered pulsars since a survey has to find them and dropping them would bias the population narrow. Each profile is a von Mises (i.e.~circular Gaussian) core with wings solved to reproduce the drawn width ratio, because a single von Mises cannot represent about a third of the measured population, all on the broad-winged side that carries less high-harmonic power. Millisecond pulsars take uniform duty cycles between 2\% and 30\%. Only a small fraction of real pulsars have extremely narrow pulses, with 12.6\% of the slow population below 1\% duty cycle, and the population-weighted results below reflect that. We oversampled the narrowest pulses and carry importance weights so that those cells have adequate statistics.

\paragraph{Brightness.} We draw the injected signal-to-noise ratio uniformly between 5.5 and 11.5, which straddles the detection threshold of every code, and is defined against a zero-mean, unit-norm template of the true pulse shape. This means that for an ``injected S/N 9'' signal, an ideal matched filter of the proper pulse shape and which removes the baseline would measure 9. This is the natural definition for a search, since every search removes the baseline, and it avoids a duty-cycle-dependent artifact of up to 30\% that a template including the mean would introduce.

\begin{table*}[t]
\centering
\caption{Detection fraction (percent) at 0.1 false alarms per observation, matched per knee-frequency bin, and (after the slash) matched per knee-frequency bin and per spin-frequency band. The two matchings are not on the same scale, since the second allows 0.1 false alarms in each of seven spin-frequency bands. ``Whitened'' rows search a time series whitened with \rednoise, with \riptide's own running-median filter turned off (\S\ref{sec:sensitivity}). Configuration C was run on white noise only without whitening, and on red noise only with it. The \rseek\ C row is from 22,368 white-noise observations and the whitened rows from 33,010 red-noise observations, where the \cs\ results agree with the full-study row to within 0.3\%. Cells marked $\dagger$ contain a substantial fraction of chance coincidences (\S\ref{sec:sensitivity}) and should not be read as sensitivities.}
\label{tab:detection}
\begin{tabular}{lcccccc}
\hline\hline
 & white & 0.1--0.5 & 0.5--2 & 2--6 & 6--15 & 15--50\,Hz \\
\hline
\cs\        & 76.3 / 84.2 & 76.2 / 84.2 & 75.2 / 82.7 & 69.8 / 77.6 & 61.3 / 68.7 & 47.9 / 54.5 \\
\rseek\ C   & 64.5 / 79.9 & --- & --- & --- & --- & --- \\
\rseek\ C, whitened & --- & 61.3 / 79.2 & 58.7 / 77.0 & 53.8 / 70.8 & 46.9 / 62.0 & 38.3 / 48.6 \\
\rseek\ B   & 71.4 / 80.7 & 69.2 / 79.6 & \phantom{0}0.1 / 71.1 & \phantom{0}0.0 / 61.6 & \phantom{0}0.1 / 45.6 & \phantom{0}0.0 / 30.9 \\
\rseek\ A   & 50.2 / 72.5 & 50.2 / 71.5 & \phantom{0}0.0 / 65.6 & \phantom{0}0.0 / 55.0$^\dagger$ & \phantom{0}0.0 / 50.8$^\dagger$ & \phantom{0}0.0 / 36.8$^\dagger$ \\
\rseek\ A, whitened & 49.0 / 71.3 & 48.7 / 71.2 & 46.9 / 69.1 & 43.5 / 62.8 & 38.9 / 55.2 & 32.0 / 44.3 \\
\accelsearch & 41.6 / 57.9 & 41.3 / 57.7 & 39.6 / 55.2 & 36.4 / 50.9 & 31.9 / 44.6 & 26.3 / 35.3 \\
\hline
\end{tabular}
\end{table*}

\begin{figure*}[t]
\centering
\includegraphics[width=0.95\textwidth]{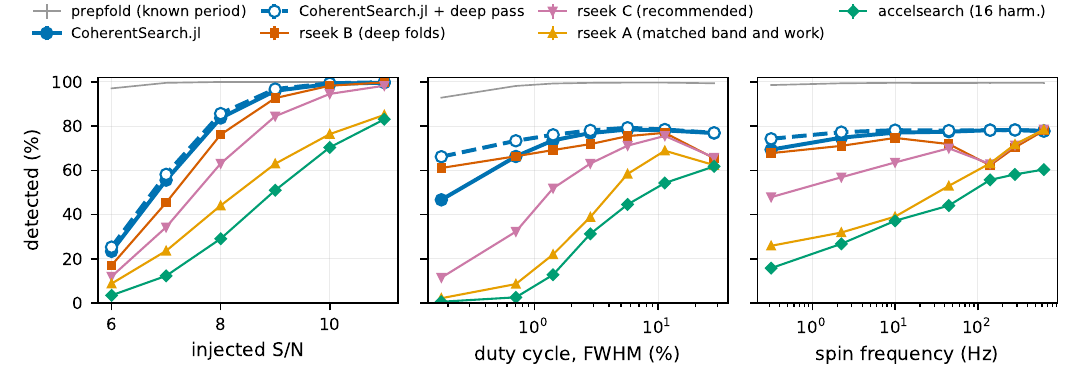}
\caption{Detection fraction in white noise against the injected signal-to-noise ratio, the injected duty cycle (full width at half maximum as a fraction of the period), and the injected spin frequency, with every code's threshold set to 0.1 false alarms per observation. ``Deep pass'' is the union of the default search with a 120-harmonic pass below 5\,Hz. The three \rseek\ configurations are described in \S\ref{sec:comparisons}. \rseek\ B ran on one observation in seven and \rseek\ C on 22,368 of the 76,105, and each is scored over the observations it ran on. \prepfold\ folds at the known period and is a best-case rather than a search competitor. Error bars are bootstrapped over observations and are mostly smaller than the symbols.}
\label{fig:white}
\end{figure*}

\paragraph{White and red noise.} Half of the injections carry only white noise. The other half add red noise, i.e.~slow wandering of the baseline of the kind that gain drift and atmospheric opacity variations produce. Its power spectrum is $1 + (\nu/\nu_{\rm knee})^{-\alpha}$, where the knee frequency $\nu_{\rm knee}$ is where the red noise falls to the level of the white noise. Fitting this form to four real dedispersed observations from Arecibo, the Green Bank Telescope, and Parkes gave $\alpha$ of 1.9--2.0 wherever the red noise mattered at all (a random walk, as expected from gain drift), with the knee spanning 1.6--31.5\,Hz. We therefore draw $\alpha$ from a Gaussian of mean 2.0 and width 0.3 and the knee log-uniformly between 0.1 and 50\,Hz, and treat the knee frequency as the axis of the study. The red observations reuse the white observations' pulsars and white noise exactly, so every red-versus-white comparison is paired on the same pulsar in the same noise. \citet{lbh+15} found that PALFA's sensitivity degradation sets in at $P \sim 100$\,ms, i.e.~a knee of order 10\,Hz, which corroborates the choice of range.

\paragraph{Scoring.} A candidate counts as a detection if it lies within half a Fourier bin of the true frequency or of a simple harmonic ratio $n/m \le 8$ of it, with harmonic detections flagged. \rseek\ does not remove harmonic families, while \cs\ and \accelsearch\ do, so counting harmonics as misses would penalize \rseek. Every candidate matching no injection is a false alarm, and we store the strongest 800 per observation for each code. \rseek's fold depth, the harmonic count of our detection, the best boxcar width, and the wall-clock time of every stage are recorded, so all of the tables below come from the same stored results.

\subsection{Results of the tests} \label{sec:results}

We are primarily interested in sensitivity, false-alarm rate, resilience to red noise, and speed. We have tried to run every code the way its authors intend, and where a choice could be made either way we have typically chosen the one which would benefit the other codes. Our goal is to compare the codes as they are intended to be used in real search pipelines.

A signal-to-noise ratio of 8 does not mean the same thing in any two of these codes. Ours, \riptide's, and \accelsearch's are all single-trial statistics, but they come from different numbers of trials, over noise of different shapes, with tails of different weight, and \prepfold's is a chi-squared test converted to an equivalent Gaussian significance. Comparing them at a common nominal threshold would be meaningless. Instead we fix the measured false-alarm rate at 0.1 false alarms per ``observation'', and let each code have whatever threshold delivers that rate. On white noise that threshold is 6.75 for \cs, 7.20 for \accelsearch, 7.55 for \rseek\ A and C, and 7.65 for \rseek\ B. A code then wins on two separable things: how low its threshold can sit, which depends on the shape of the tail of its noise distribution, and how much signal it measures above that threshold, which is how well its filter matches the pulse.

Under red noise there is an additional complication. The false-alarm rate of some codes depends on the red-noise level and on the trial frequency. For those codes, one threshold chosen to give 0.1 false alarms per observation on average over all of the red-noise levels would give far more than that in the worst observations and fewer in the mildest, so it would not describe any of them. We therefore match the rate within bins of knee frequency, and separately within bins of both knee frequency and the pulsar's own spin frequency, which is the treatment that gives \riptide\ its best performance (\S\ref{sec:sensitivity}).

\begin{figure}[t]
\centering
\includegraphics[width=\linewidth]{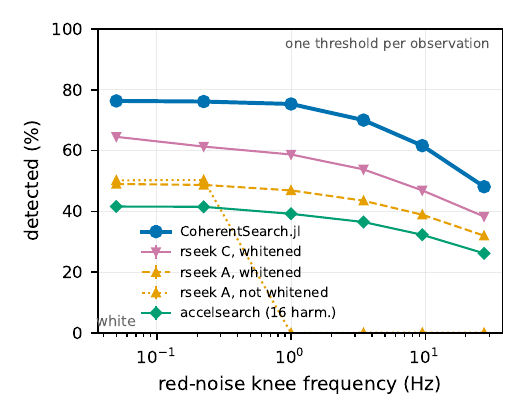}
\caption{Detection fraction against the knee frequency of the injected red noise, with the leftmost point the white-noise observations. Each code has one threshold per observation, matched to 0.1 false alarms per observation within each knee-frequency bin, as a search pipeline would apply it. The whitened \rseek\ curves search a time series whitened with \rednoise\ and have \riptide's own running-median filter turned off (\S\ref{sec:sensitivity}). Their leftmost points are \rseek\ C and A on unwhitened white noise, where whitening changes the result by about 1\%. Without whitening, \rseek's slow trials flood its candidate list above a knee of 0.5\,Hz and its detection fraction falls to zero. Matching the threshold separately in each spin-frequency band instead raises every curve (Table~\ref{tab:detection}, second numbers). The red-noise points are from the 33,010 observations on which every code ran.}
\label{fig:red}
\end{figure}

\subsubsection{Sensitivity} \label{sec:sensitivity}

\paragraph{White noise.}
On data containing white noise only, \cs\ recovers 76.3\% of the injected pulsars. \rseek\ recovers 64.5\% in configuration C, its recommended regime, which takes about three times our run time. It recovers 50.2\% in configuration A, matched to our coverage and our amount of work, and 71.4\% in configuration B, at about seven times our run time. \accelsearch\ recovers 41.6\%, although it does so much more quickly. \prepfold, told the period, reaches 99.4\%. Table~\ref{tab:detection} gives these numbers, along with the red-noise results discussed below.

Figure~\ref{fig:white} shows how the detection fraction depends on the injected brightness, pulse width, and spin frequency. The overall fraction hides where each code performs better or worse, and duty cycle is the axis that matters most. Narrow pulses are the hardest case for any search, and they are where fast folding earned its reputation \citep{kml+09,cbc+17,mbs+20}: a folding search keeps the pulse shape at any depth it is given, where an FFT search summing as many as 16 or 32 harmonic powers both discards the shape and leaves signal at unsummed harmonics. In configuration A, \rseek\ folds no deeper than we do and finds far fewer of these pulsars: 8.6\% against our 66.3\% at duty cycles of 0.5--1\%, and 2.2\% against 46.7\% below that. Configuration C sits between A and B at every duty cycle. Its 120--140-bin folds roughly double the detections of configuration A between 1\% and 4\% duty cycle, and it matches configuration B above 8\%. Below 1\% it finds 11.5\% and 32.3\% of the pulsars (below 0.5\%, and from 0.5\% to 1\%) against our 46.7\% and 66.3\%, because its folds stop at 140 bins. In configuration B, though, \rseek\ folds 960--1040 bins where we fold 120, and it beats our default configuration below 0.5\% duty cycle, 61.1\% against 46.7\%, and ties it between 0.5\% and 1\%. Above 1\% our default is ahead of all three configurations at every width, and it is basically flat in spin frequency where \rseek\ A and \accelsearch\ climb toward the fast end. Our loss at the narrowest duty cycle is harmonic truncation, not the boxcar filter. A model of the recovered statistic under our bank of boxcar templates agrees with an ideal matched filter truncated at $H$ harmonics to 0.1\% at these widths, and at 0.3\% duty cycle a pulse spans less than half of one of our 120 profile bins.

However, depth is fairly inexpensive in the Fourier domain, so we can afford a deeper fold where it is needed. A second pass with 120 harmonics and decimation factors up to 12, over spin frequencies below 5\,Hz only, covers every injected pulsar below 0.5\% duty cycle (all of which are slower than 1.7\,Hz in this population) and takes 3.1\,s per observation against the default's 10.8\,s on one CPU core (\S\ref{sec:speed}). On its own it finds 71.3\% of the narrowest pulsars, the best of any search in the study. The union of the two passes, with its threshold measured on the concatenated candidate lists so that the second pass pays for its own trials, is ahead of \rseek\ B at every duty cycle (Figure~\ref{fig:white}, middle), from 66.2\% against 61.1\% at the narrowest to 77.0\% against 65.5\% at the widest, for 13.9\,s per observation for \cs\ against 79.1\,s for \rseek. Over the whole population the second pass adds only 1.3 points, because the pulsars it benefits are a small slice of the population, and the default configuration alone remains the subject of this paper. What the deep pass shows is that a Fourier-domain search can reach the depth a very narrow pulse needs for a computational cost a survey can likely afford. \riptide's deep configuration is doing exactly what it should, and a fifth of its cost in the Fourier domain matches or even improves on it.

Fitting a logistic curve to the detection fraction against injected signal-to-noise ratio (Figure~\ref{fig:white}, left) gives the brightness at which each code detects half of the injections. Over all duty cycles it is 6.84 for \cs, 7.15 for \rseek\ B, 7.58 for \rseek\ C, 8.49 for \rseek\ A, and 9.07 for \accelsearch. Below 1\% duty cycle it is 7.64 for our default, 7.57 for \rseek\ B, 9.86 for \rseek\ C, and 6.77 for our deep pass.

\paragraph{Red noise.}
Red noise is a real problem in modern pulsar searches, and in these simulations it can effectively break \riptide\ due to the default time-domain de-trending method used. \riptide\ removes slow red-noise-like drifts by subtracting a running median from the time series, with a default window of 4\,s. That is effectively a high-pass filter at $1/({\rm window})$ which cannot remove wander faster than about 0.25\,Hz. PRESTO's \rednoise\ removes the wander in the frequency domain instead, across the whole spectrum. So when the knee frequency of the noise rises above $\sim$0.3\,Hz, \rseek's candidate list fills with false positives at slow periods. Its threshold for 0.1 false alarms per observation climbs from 7.6 to 250 across our tested red noise knee range, and one threshold for the whole observation, which would likely be used in practice, causes its detection fraction to effectively go to zero since it is swamped by false positives (Table~\ref{tab:detection}, first number in each cell, and Figure~\ref{fig:red}). \accelsearch's threshold is flat because of its own local normalization of the power spectrum, and ours is flat because the input FFT is whitened. This is \riptide\ running the way its authors intend, and it is not a defect in its search.

Giving \riptide\ its own threshold in each spin-frequency band is a fairer test of its search, since its fast folds stay clean while its slow ones are flooded by false positives. Under that treatment (Table~\ref{tab:detection}, second number in each cell) \rseek\ B recovers to 79.6\% at the mildest knee and 30.9\% at the most severe, against our 84.2\% and 54.5\%. Both numbers in Table~\ref{tab:detection} are important since per-knee matching is what a pipeline with one threshold per observation applies, and per-band matching is what \riptide's search can do if its statistic is calibrated. Band matching cannot fully repair a flooded candidate list, however. Some of the false positives land within the scoring tolerance of an injection by chance and count as detections, which a matched false-alarm rate cannot see. We estimate this from the frequency offsets of the claimed candidates. Above a 2\,Hz knee it is a substantial fraction of the band-matched detections of \rseek\ A, at least 16\% at the worst knee against 0.27\% of ours, so we do not quote those cells as sensitivities.

The fairest test of \riptide\ under red noise gives it the same whitened data that \cs\ searches. We transformed each \rednoise-normalized FFT back into a time series and ran \rseek\ on it with its running-median filter turned off, since leaving it on would filter the data a second time. \riptide's own normalization of the time series is kept. Its threshold for 0.1 false alarms per observation is then the same in every knee bin, 7.55 for configuration A and 7.55--7.65 for configuration C, so a single threshold per observation works again, and chance coincidences fall to 0.1--0.2\% of its detections. In configuration C it recovers 61.3\% of the pulsars at the mildest knee and 38.3\% at the most severe, against 76.1\% and 48.1\% for \cs\ on the same observations (Table~\ref{tab:detection} and Figure~\ref{fig:red}). Both codes keep 63\% of their mild-noise sensitivity at the worst knee, and whitening costs \riptide\ about 1\% on white noise (49.0\% against 50.2\% for configuration A). So with the same whitening, the FFA and \cs\ lose sensitivity to red noise in the same proportion, and the collapse of \rseek\ under red noise comes entirely from its de-trending step. We did not run configuration C on red noise without whitening, but it uses the same running median as configurations A and B, and would very likely fail the same way.

Red noise costs \cs\ little, and only at low frequencies. The injected brightness we need for 50\% detection rises by factors of 1.00, 1.00, 1.05, 1.12, and 1.27 as the knee rises through the five bins of Table~\ref{tab:detection}, and the paired red-minus-white difference in our statistic is exactly zero above 100\,Hz at every knee. \citet{lbh+15} injected pulsars into real PALFA observations and measured a degradation of 1.1--2 in the minimum detectable flux for $P = 0.1$--2\,s at ${\rm DM} > 150$~pc~cm$^{-3}$, and up to 10 for $P > 4$\,s at low DM. We sit at or below the bottom of their range, which makes sense as their observations contain radio-frequency interference as well as red noise, and we model only the latter.

Whitening the input, whether time series or Fourier spectrum, is not optional. Running \cs\ on the raw FFT, with $\sigma$ measured from the data since the analytic value assumes a normalized input, drops the detection fraction from 77\% to 1\% across the same knee range, and the matched threshold has to rise from 6.8 to 11.8 to hold the false-alarm rate. A single scale factor cannot fix a red spectrum since a trial at $r$ sums harmonics out to $60r$, where we may be mixing bins whose noise powers differ by orders of magnitude, with the red bins dominating. Fourier-domain search techniques need the FFT normalized locally, so that the mean noise power is one (or at least known, for statistical calculations) at every frequency. PRESTO's \rednoise\ is one way to do this, but definitely not the only way. With a normalized spectrum, measuring $\sigma$ from the whitened data instead of computing it changes the detection fraction by less than a point for every knee bin, so the free, and more stable, analytic value is the one to use.

\subsubsection{False-alarm rate} \label{sec:fap}

The threshold for \cs\ sits lower than the others' for the same false-alarm rate (6.75 against 7.2 for \accelsearch\ and 7.55--7.65 for \rseek\ on white noise), which is where most of the improvement in sensitivity comes from. Configurations A and C of \rseek\ need the same threshold, 7.55. Scored at that threshold instead of its own, \cs\ would detect 61.6\% of the white-noise injections, more than the 50.2\% of configuration A but less than the 64.5\% of configuration C. Against configuration A, then, a little over half of our lead comes from the lower threshold and the rest from recovering more signal. Against configuration C the lower threshold accounts for all of our lead, and C's folds recover slightly more signal than ours overall. For pulsars detected by both, \cs\ recovers more signal than \rseek\ C below 2\% duty cycle and less above 4\%, which is the expected pattern for a 120--140-bin fold against our ladder of fold depths.

Three things keep our noise tail well behaved, with the most important being that the noise scale is computed analytically from the normalized input rather than estimated from each profile. This means there is no per-profile scatter in the statistic. On injections of fixed brightness the standard deviation of our recovered statistic is 1.0, against 1.6 for \rseek\ and 1.9 for \prepfold, whose estimators of $\sigma$ from a few dozen or hundred bins add their own noise. That noise inflates both the recovered signal and the false-alarm tail, so at a matched false-alarm rate it turns into a loss. Second, the boxcar variance of Equation~(\ref{eq:varsw}) puts every fold depth and every frequency on a common ground, so one threshold serves the whole search. And third, our false-alarm tail is Gaussian to the digit, falling by the expected factor of $\exp[-(7^2 - 6^2)/2]$ between S/N thresholds of 6 and 7, where \rseek's is heavier (i.e.~more populated) than Gaussian by a factor of about 8 over the same span. Additionally, the harmonic collapse of \S\ref{sec:details} keeps the candidate list short and more accurate.  \cs\ reports a median of 12 candidates per observation where \rseek\ A reports about 160. On white noise 97.7\% of our matched candidates are at the fundamental rather than a harmonic, against 92.7\% for \rseek\ A and 93.6\% for \accelsearch. \rseek\ C and B do better than all three, at 98.7\% and 98.9\%, likely because their deeper slow folds less often detect a slow pulsar at a multiple of its frequency.

For \riptide\ the false-alarm rate depends on the red-noise level and on the trial period, because its running-median filter leaves the slow trials contaminated. A survey using it therefore needs a threshold per frequency band and per observation, or a dynamic threshold like the one its authors describe \citep{mbs+20}. It is important to reiterate that the primary reason for this is the time-domain detrending and not the FFA algorithm, in general.

\subsubsection{Speed (single-threaded)} \label{sec:speed}

\begin{figure}[htb]
\centering
\includegraphics[width=\linewidth]{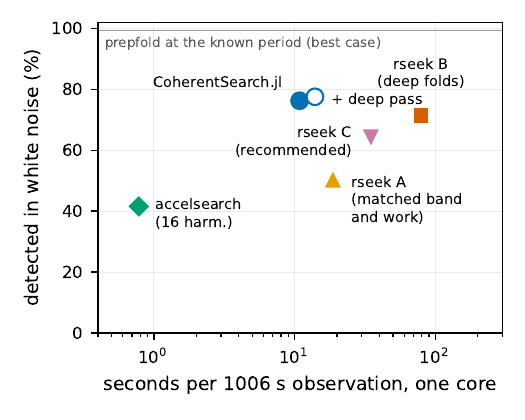}
\caption{Sensitivity against cost. The white-noise detection fraction of each search at 0.1 false alarms per observation, against its median wall-clock time per 1006\,s simulated observation, measured for every search on the same 49 observations as one single-threaded process on an otherwise idle Xeon w5-3433. The FFT and normalization that the Fourier-domain searches share (under 1\,s) are not included. The horizontal line is \prepfold\ folding at the known period. The deep pass adds 3.1\,s to the default search.}
\label{fig:cost}
\end{figure}

Figure~\ref{fig:cost} plots each search's white-noise detection fraction against its median wall-clock time per simulated observation. We measured these times by running every search back to back, one at a time and single-threaded, on the same 49 simulated observations on an otherwise idle 16-core workstation (a Xeon w5-3433). The spread between observations is under 1\% for every search except \accelsearch, whose sub-second times vary by about 5\%. \cs\ takes 10.8\,s per observation. \rseek\ takes 1.7 times as long in configuration A, 3.2 times in configuration C, and 7.3 times in configuration B, while detecting a smaller fraction of the pulsars in each case. \accelsearch\ is 14 times faster than \cs, and detects 41.6\% of the pulsars compared with our 76.3\%. The same measurement for configurations A and C on a different workstation (two Xeon Silver 4114 processors) gave ratios of 1.7 and 3.1, so the ratios depend little on the machine. Timing the codes while many simulations ran in parallel on the same machine, as during the Monte Carlo itself, inflates every time by different amounts, so we do not use those times for comparisons. A direct head-to-head on a real Parkes observation (a 2097\,s pointing, both codes covering 0.1--200\,Hz single-threaded, medians of three runs) shows \cs\ being 1.4--2.2 times faster than \rseek\ on three different CPUs, with the ratio depending more on the host than on anything in the code. This is the case even though \cs\ folds all frequencies multiple times, compared to once for \rseek, due to the decimation ladder of \S\ref{sec:decimation}.

Start-up costs of the code can be significant and need to be addressed. Julia compiles code on first use, and in earlier versions of the code, that compilation cost could dominate the cost of a search by factors of several. Rearranging the code to allow much more pre-compilation in the package image and keeping the command-line driver inside the package cut that to 2.4\,s, of which Julia's own start is 0.2\,s. On a GPU, loading the CUDA runtime adds about 6\,s once. However, these start-up costs only occur once per invocation, so the command line accepts many FFT files and shares its interpolation tables, FFT plans, and per-thread workspaces across them, meaning that the marginal cost of an extra file is about one second plus its own search. A real survey should pass a whole range of DM trials to one invocation rather than start the program once per file.

How close is the single-threaded code to as fast as it can go? Searching the Parkes observation above over 0.1--33.3\,Hz on one core of a Xeon Silver 4114 workstation, the batched inverse FFTs take 44\% of the run time, the boxcar matched-filter scan 30\%, and the Fourier interpolation 24\%. The interpolation already uses SIMD instructions across trials and reads each FFT bin from cache. The inverse FFTs of the decimated folds read every $k$th harmonic from memory rather than a contiguous block, which makes them about 15\% slower than the same transforms on contiguous data. That is at most 4\% of the total run time, and the fastest way we have found to copy the harmonics into contiguous memory first costs more than twice what it saves. Every change we have tried to the boxcar scan, such as merging it with the steps around it, has made it slower. So we do not expect further tuning of the CPU code to gain more than a few percent.

\subsubsection{Parallelism, and how real searches would be run} \label{sec:parallel}

\begin{figure}[htb]
\centering
\includegraphics[width=\linewidth]{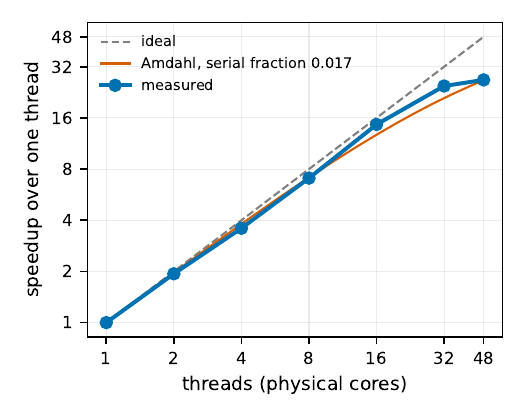}
\caption{Wall-clock speedup of the warm (i.e.~in cache), in-process search against thread count on two AMD EPYC 7413 processors (48 physical cores), searching a 7.3-hour Green Bank Telescope observation over 0.1--33.3\,Hz. The dashed line is ideal scaling and the thin solid line an Amdahl's law fit with a serial fraction of 0.017.}
\label{fig:scaling}
\end{figure}

The frequency trials are independent, so the search parallelizes in three ways, and which to use depends on the type of search desired. Within one invocation, groups of trials are dealt to threads round-robin, each with a private workspace. Figure~\ref{fig:scaling} shows the speedup on a 48-core machine, which achieves 26.8 times speed-up with 48 threads, with 88--97\% efficiency through 16 threads. A fit of Amdahl's law gives a serial fraction (i.e.~the part of the work that cannot run in parallel) of 0.017. On three of the four machines where we tried more threads than physical cores, the search got slower, by up to a factor of two. On the fourth, a 20-core Xeon workstation, 40 threads were about 7\% faster than 20 for a long search. We therefore recommend one thread per physical core unless a quick test on a given machine shows otherwise. We note that \riptide's C extension is built without OpenMP or explicit multi-threading, so \rseek\ cannot use multiple cores.

A blind survey with unknown DM usually has more time series than cores, and there the right model is parallelism over DM trials using single-threaded processes. The \cs\ repository contains a helper script called {\tt parallel\_search.py} which partitions the DMs across multiple \cs\ jobs, resulting in the start-up costs being paid only once. When there is only a single time series to search, such as when a DM is known {\em a priori}, as for a globular cluster or a search for the companion of a known pulsar after acceleration correction, multi-threading is the best way to use a whole machine.

Finally, the whole search can also run on an NVIDIA GPU, through the {\tt CUDA.jl} package \citep{cudajl18} that a CPU-only user never downloads or pre-compiles. Table~\ref{tab:gpu} lists the six cards we have measured on one large observation over 0.1--133\,Hz, 423 million trial fundamentals. We compare each card with a whole 20-core Xeon workstation, i.e.~both of its processors with all 20 cores searching together, one thread per core. An L40 or an A100 is nine times faster than all 20 of those cores, an RTX A4000 three times, and a small RTX A400 is slower than them. All six report the same 285 candidates as the CPU, differing at most in the last digit of a printed period. A machine with many CPU cores is competitive with a mid-range card: the 48-core machine of Figure~\ref{fig:scaling} is faster than its own A4000, and a 32-core Threadripper is 6.7 times faster than the low-end workstation A400 GPU beside it. The GPU code is single precision throughout and agrees with the CPU to $\sim 2 \times 10^{-7}$ on generated profiles and on the statistic.

\begin{table}[t]
\centering
\caption{GPU search of a 7.3-hour observation (1.29\,GB FFT) over 0.1--133.3\,Hz, 423 million trial fundamentals with decimation to 800\,Hz, at each card's best trial-group size. The CPU reference is the same search on both Xeon Silver 4114 processors of one workstation, 20 cores in total, with one thread per core (40.1\,s). Using all 40 hardware threads of that machine is 7\% faster (37.5\,s).}
\label{tab:gpu}
\begin{tabular}{lccc}
\hline\hline
 & & & speedup vs. \\
Card & wall (s) & ns per trial & 20 CPU cores \\
\hline
L40           & 4.30  & 10.2  & 9.32 \\
A100 (80\,GB)  & 4.52  & 10.7  & 8.88 \\
RTX 4500 Ada  & 7.30  & 17.2  & 5.49 \\
RTX A4000     & 12.07 & 28.5  & 3.32 \\
GTX 1080      & 29.9  & 70.7  & 1.34 \\
RTX A400      & 61.4  & 145   & 0.65 \\
\hline
\end{tabular}
\end{table}

\subsubsection{Comparison with other papers comparing codes} \label{sec:otherpapers}

Several studies have compared FFT and FFA searches, or measured what red noise and its removal cost an FFT pipeline, and our results are consistent with them where they overlap. \citet{vkr17} built a framework for assessing pipelines against non-stationary noise and weak interference, and tested SIGPROC and PRESTO with 15 injected pulsars of fixed 12\% duty cycle at one noise realization per cell. They found that spectral whitening (SIGPROC's block normalization and PRESTO's \rednoise) controls false positives, at 20 per true positive against 118 to 900 without it, but attenuates the fundamentals of slow pulsars along with the noise. We see the same trade-off in our own red-noise results. Whitening keeps our threshold flat as a function of spin frequency, and its cost is confined to the slow end, a factor of 1.27 in sensitivity at the worst (i.e.~highest frequency) knee for pulsars below 1\,Hz and nothing above 100\,Hz. \citet{lbh+15} measured the net effect on PALFA with flux-calibrated injections into real observations, and we sit at the low end of their degradation factors, as discussed above.

\citet{srp+22} compared \riptide\ with \accelsearch\ on real GHRSS noise, injecting one pulsar per parameter setting at fixed brightness and reporting the recovered signal-to-noise ratio normalized to its maximum, with \accelsearch\ summing 8 harmonics. They found the FFA's recovered signal flat in period and duty cycle where the FFT's fell by more than 80\% by $P = 10$\,s in real noise (going to longer periods), and that running-median subtraction before the FFA outperformed \rednoise\ before either search. Their FFT result is what an 8-harmonic incoherent sum on a whitened spectrum does to a slow pulsar, and our \accelsearch\ column agrees with its direction. Their conclusion about the two de-reddening methods is about which is better for a \emph{time-domain} search, and is not in tension with ours: a Fourier-domain search needs the spectrum normalized, and once it is, our threshold does not move. \citet{gbm24} ran 1000 white-noise realizations per cell at fixed 4\% duty cycle, varying period, nulling, and brightness, with \accelsearch\ summing 32 harmonics, and found the FFA and FFT comparable, the FFA slightly ahead for the faintest and most strongly nulling signals, and \riptide's pipeline one to two orders of magnitude slower. That is close to our white-noise picture at 4\% duty cycle, where \rseek\ B and \accelsearch\ at 32 harmonics would be expected to sit between the curves of Figure~\ref{fig:white}.

The present study differs from these in scale and in method: 869,574 injections rather than tens or thousands; duty cycles and profile shapes drawn from a measured population rather than fixed; white and red noise in paired realizations; and detection fractions at a matched empirical false-alarm rate rather than recovered signal-to-noise ratios at nominal thresholds. It does not include radio-frequency interference, deliberately. \riptide\ has no interference-excision stage, since it detrends in the time domain, so zapping the spectrum would be a step we imposed on it, and \citet{vkr17} found that the PRESTO masking they tested had little effect on sensitivity in any case.

\section{Conclusions} \label{sec:conclusions}

The two families of pulsar search code in common use each have substantial weaknesses. The incoherent harmonic sum of an FFT search is fast, but it discards the pulse shape and, at 16 or fewer harmonics, most of the power of a narrow pulse. In our injection study \accelsearch\ recovers 42\% of a realistic population at a false-alarm rate of 0.1 per observation. The FFA keeps the pulse shape and, given enough phase bins, is the more sensitive of the two, but it is much slower, and its time-domain de-reddening leaves its false-alarm rate dependent on the red-noise level and on the trial period, so that a single threshold per observation fails under red noise and its candidate lists are long. Given a time series whitened in the Fourier domain instead, its false-alarm rate is as stable as ours and it loses sensitivity to red noise in the same proportion.

\cs\ takes the FFT search's input, one long transform, and does the FFA's job in the Fourier domain. It interpolates the complex amplitudes of 60 harmonics at each trial frequency, inverts them into a pulse profile, and matched filters that profile against boxcars, reusing the same harmonics to fold at two to six times each trial frequency almost for free. Because the input spectrum is normalized, the noise in every profile is known in closed form, and one threshold gives one false-alarm rate across all fold depths and all frequencies. On the same injections it recovers 76\% of the population, against 64\% for \riptide's FFA in its recommended configuration at about three times the cost, 71\% for its deepest configuration at about seven times the cost, and 50\% for the configuration matched to our coverage and work. Against the recommended configuration all of that margin comes from the lower threshold that our calculable noise allows at the same false-alarm rate. \cs's detection threshold does not move under red noise, its candidate lists are an order of magnitude shorter, and it is faster than \rseek\ single-threaded, scales across all the cores of a machine, and runs on a GPU. The one place its default configuration loses is to a much deeper FFA fold at duty cycles below 0.5\%, and a second pass at twice the depth over the slowest frequencies makes it the best performing search for a fifth of the FFA's cost. It is slower than \accelsearch\ by an order of magnitude, and a survey that wants to skim the brightest pulsars first should still run \accelsearch\ or another incoherent FFT search first.

The method is not specific to pulsars. Any search for a periodic signal whose shape is narrow compared with its period, and whose period is unknown, is a search over harmonic phases as well as amplitudes. Exoplanet transit searches are one example. A transit is a box-shaped dip of a few percent of the period. The standard box least-squares algorithm \citep{kzm02} folds the light curve at each trial period and fits a box to it, and the transit least-squares algorithm \citep{hh19} replaces the box with a limb-darkened transit shape. That is what Figure~\ref{fig:pseudocode} does in the Fourier domain. Adapting the code would need an FFT normalization suited to stellar variability and a bank of dips rather than pulses, but nothing in the search itself would change.

One thing the Fourier domain does not easily give us is an acceleration search \citep[e.g.~][]{rem02}. A pulsar in a tight binary changes its apparent spin frequency during an observation, and its harmonics smear over neighboring bins in proportion to the harmonic number. The efficiencies described above, which make \cs\ quite fast given its sensitivity, especially the tabulated interpolation weights and the reuse of harmonics across decimations, rely on every harmonic being a fixed multiple of one trial frequency, and a Fourier-domain acceleration search would break that. For accelerated pulsars the practical route would be a time-domain acceleration search \citep[e.g.~][]{mk84,jk91}.  Such searches stretch or compress the time series for each trial acceleration before the FFT, and would be fully compatible with \cs, just as they are for an FFA-based search \citep{wcb+24}.

\begin{acknowledgments}
The National Radio Astronomy Observatory is a facility of the National Science Foundation operated under cooperative agreement by Associated Universities, Inc. SMR is a CIFAR Fellow and is supported by the NSF Physics Frontiers Center award 2020265, NSF Grant 2607948, and NSF AAG award 2510064. We thank Vincent Morello for his comments on a draft of this paper and for suggesting configuration C of \riptide. Thanks also go to Nolan Denman, Wolfgang Baudler, Dave MacMahon, and John Tobin for making workstations and GPUs available for testing.
\end{acknowledgments}

\begin{contribution}
SMR conceived the algorithm, wrote the Python reference implementation, designed the comparisons, wrote a detailed paper outline, and edited the manuscript. The Julia implementation, its optimization and GPU port, the Monte Carlo driver and analysis, and the first draft of this manuscript were written by Claude \citep[][primarily Opus 5 series for coding and Fable 5.1 for manuscript draft]{claude26} under SMR's direction, with every numerical result verified against the Python reference and the reference paths, including real telescope data, in the test suite.
\end{contribution}


\software{PRESTO \citep{presto_ascl},
          riptide \citep{riptide_ascl},
          Julia \citep{julia17},
          FFTW \citep{fj05},
          CUDA.jl \citep{cudajl18},
          NumPy \citep{hmv+20}
          }

\bibliography{ransom_coh_search}
\bibliographystyle{aasjournalv7.1}



\end{document}